\documentclass{article}
\usepackage{spconf,amsmath,graphicx,bbm}
\usepackage{multirow, hhline, stmaryrd}
\DeclareMathOperator*{\argmax}{arg\,max}
\DeclareMathOperator*{\argmin}{arg\,min}

\title{Speaker-Invariant Training via Adversarial Learning}
\name{Zhong Meng$^{1,2}$\sthanks{Zhong Meng performed the work while he was a research intern
	at Microsoft AI and Research, Redmond, WA, USA.}, Jinyu Li$^{1}$,
	Zhuo Chen$^{1}$, Yong Zhao$^{1}$, Vadim Mazalov$^{1}$, Yifan Gong$^{1}$,}
\secondlinename{Biing-Hwang (Fred) Juang$^{2}$}

\address{$^{1}$ Microsoft AI and Research, Redmond, WA, USA
\\ $^{2}$ Georgia Institute of Technology, Atlanta, GA, USA
}

\begin{document}
\ninept
\maketitle

\begin{abstract}

We propose a novel adversarial multi-task learning scheme, aiming at actively curtailing the inter-talker feature variability while maximizing its senone discriminability so as to enhance the performance of a deep neural network (DNN) based ASR system. We call the scheme speaker-invariant training (SIT). In SIT, a DNN acoustic model and a speaker classifier network are jointly optimized to minimize the senone (tied triphone state) classification loss, and simultaneously mini-maximize the speaker classification loss. A speaker-invariant and senone-discriminative deep feature is learned through this adversarial multi-task learning. With SIT, a canonical DNN acoustic model with significantly reduced variance in its output probabilities is learned with no explicit speaker-independent (SI) transformations or speaker-specific representations used in
training or testing. Evaluated on the CHiME-3 dataset, the SIT achieves 4.99\% relative word error rate (WER) improvement over the conventional SI acoustic model. With additional unsupervised speaker adaptation, the speaker-adapted (SA) SIT model achieves 4.86\% relative WER gain over the SA SI acoustic model.


\end{abstract}

\section{Introduction}
\label{sec:intro}
The deep neural network (DNN) based acoustic models have been widely used
in automatic speech recognition (ASR) and have achieved extraordinary
performance improvement \cite{DNN4ASR-hinton2012,  yu2017recent}. 
However, the performance of a speaker-independent (SI) acoustic model
trained with speech data from a large number of speakers is still 
affected by the spectral variations in each speech unit caused by the 
inter-speaker variability. Therefore, speaker adaptation methods are widely used to
boost the recognition system performance \cite{saon2013speaker, svd_xue_2, xue2014fast, miao2015speaker, wu2015multi, svd_zhao, map_huang, multi_huang, lhuc_pawel_1, tan2016cluster, smarakoon2016factorized}. 

Recently, adversarial learning has captured great attention of deep
learning community given its remarkable success in estimating generative
models \cite{gan}. In speech, it has been applied to noise-robust 
\cite{grl_shinohara, grl_serdyuk, grl_sun, dsn_meng, meng2018adversarial} and conversational
 ASR \cite{saon2017english} using gradient reversal layer
\cite{grl_ganin} or domain separation network \cite{dsn}. Inspired by this,
we propose \emph{speaker-invariant training (SIT)} via adversarial
learning to reduce the effect of speaker variability in acoustic modeling.
In SIT, a DNN acoustic model and a DNN speaker classifier are jointly
trained to simultaneously optimize the primary task of minimizing the
senone classification loss and the secondary task of mini-maximizing the
speaker classification loss. Through this adversarial multi-task learning
procedure, a feature extractor is learned as the bottom layers of the DNN
acoustic model that maps the input speech frames from different speakers
into \emph{speaker-invariant} and senone-discriminative deep hidden features,
so that further senone classification is based on representations
with the speaker factor already normalized out.
The DNN acoustic model with SIT can be directly used to generate word
transcription for unseen test speakers through \emph{one-pass online} 
decoding. On top of the SIT DNN,
further adaptation can be performed to adjust the model towards the test speakers, 
achieving even higher ASR accuracy. 

We evaluate SIT with ASR experiments on CHiME-3 dataset, the SIT DNN
acoustic model achieves 4.99\% relative WER improvement over the baseline
SI DNN.  Further, SIT achieves 4.86\% relative WER gain over the SI
DNN when the same unsupervised speaker adaptation process is performed on both
models. With t-distributed stochastic neighbor 
embedding (t-SNE) \cite{maaten2008visualizing} visualization, we show that, 
after SIT, the deep feature distributions of different speakers are well aligned 
with each other, which demonstrates the strong capability of SIT in reducing 
speaker-variability.


\section{Related Work}
\label{sec:relate}


Speaker-adaptive training (SAT) is proposed to generate canonical 
acoustic models coupled with  speaker adaptation.
For Gaussian mixture
model (GMM)-hidden Markov model (HMM) acoustic model, SAT applies
unconstrained \cite{anastasakos1996compact} or constrained \cite{gales1998maximum}
model-space linear transformations that separately model the
speaker-specific characteristics and are jointly estimated with the GMM-HMM
parameters to maximize the likelihood of the training data. Cluster-adaptive training (CAT) \cite{gales2000cluster} is then proposed to use a linear interpolation of all
the cluster means as the mean of the particular speaker instead of a single cluster as
representative of a particular speaker. However, SAT 
of GMM-HMM needs 
to have two sets of models, the SI model and canonical model. During testing, the SI model 
is used 
to generate the first pass decoding transcription, and the canonical model is 
combined with speaker-specific transformation to adapt to the new speaker.

For DNN-HMM acoustic model, CAT \cite{tan2016cluster} and 
multi-basis adaptive neural networks \cite{wu2015multi}
are proposed to represent 
the weight and/or the bias of the speaker-dependent (SD) affine transformation 
in each hidden layer of a DNN acoustic model as a linear combination of SI
bases, where the combination weights are low-dimensional SD speaker representations. 
The canonical SI bases with reduced variances are jointly optimized with the SD
speaker representations during the SAT to minimize the cross-entropy
loss. During unsupervised adaptation, the test speaker representations are  
re-estimated using alignments from the first-pass decoding of the test data with 
SI DNN as the supervisions and are used in the second-pass decoding to generate the 
transcription. Factorized hidden layer \cite{smarakoon2016factorized} is similar to
\cite{tan2016cluster, wu2015multi}, but includes SI DNN weights as part of the linear combination.
In \cite{xue2014fast}, 
SD speaker codes are transformed by a set of SI matrices and then directly 
added to the biases of the hidden-layer affine transformations. The speaker codes
and SI transformations are jointly estimated during SAT. 
For these methods,  
two passes of decoding are required to generate the final transcription in unsupervised 
adaption setup, which increases the computational complexity of the system.


In \cite{miao2015speaker, saon2013speaker}, an SI adaptation network is learned to derive
speaker-normalized features from i-vectors to train the canonical DNN acoustic model. The i-vectors for the test speakers are then estimated and used for decoding after going through the SI adaptation network. In \cite{saon2017english}, a reconstruction network is trained to predict the input i-vector given the speech feature and its corresponding i-vector are at the input of the acoustic model. The mean-squared error loss of the i-vector reconstruction and the cross-entropy loss of the DNN acoustic model are jointly optimized through adversarial multi-task learning. Although
these methods generate the final transcription with one-pass of decoding, 
they need to go through the entire test utterances in order to estimate the i-vectors,
making it impossible to perform online decoding.
Moreover, the accuracy of i-vectors estimation are limited by the duration of the test utterances. The 
estimation of i-vector for each utterance also increases the computational complexity of the system.

SIT directly minimizes the speaker variations by optimizing an adversarial multi-task objective 
other than the most basic cross entropy object as in SAT. It forgoes the need of estimating
any additional SI bases or speaker representations during training or
testing. The direct use of SIT DNN acoustic model in testing enables the generation 
of word transcription for unseen test speakers through \emph{one-pass online} decoding. 
Moreover, it effectively suppresses the inter-speaker variability via a lightweight system 
with much reduced training parameters and computational complexity.
To achieve additional gain, unsupervised speaker adaptation can also be further conducted 
on the SIT model with one extra pass of decoding.

\section{Speaker-Invariant Training}
\label{sec:sit}

To perform SIT, we need a sequence of speech
frames $X=\{x_{1}, \ldots, x_{N}\}$, a sequence
of senone labels $Y=\{y_{1}, \ldots, y_{N}\}$ aligned with $X$ and a
sequence of speaker labels $S=\{s_{1}, \ldots, s_{N}\}$ aligned with
$X$. 
The goal of SIT is to reduce the variances of hidden and output units
distributions of the DNN acoustic model that are caused by the inherent
inter-speaker variability in the speech signal.
To achieve speaker-robustness, we learn a
\emph{speaker-invariant} and \emph{senone-discriminative} deep hidden feature in
the DNN acoustic model through adversarial multi-task learning and make
senone posterior predictions based on the learned deep feature. In order to do so, we
view the first few layers of the 
acoustic model as a
feature extractor network $M_f$
with parameters $\theta_f$ that maps 
$X$ from different speakers to deep hidden features
$F=\{f_1, \ldots, f_N\}$ (see Fig. \ref{fig:sit}) and the upper layers of the acoustic model as a
senone classifier $M_y$ with parameters $\theta_y$ that maps the intermediate
features $F$ to the senone posteriors $p_y(q|f_i; \theta_y), q\in
\mathcal{Q}$ as follows:
\begin{figure}[htpb!]
	\centering
	\includegraphics[width=0.9\columnwidth]{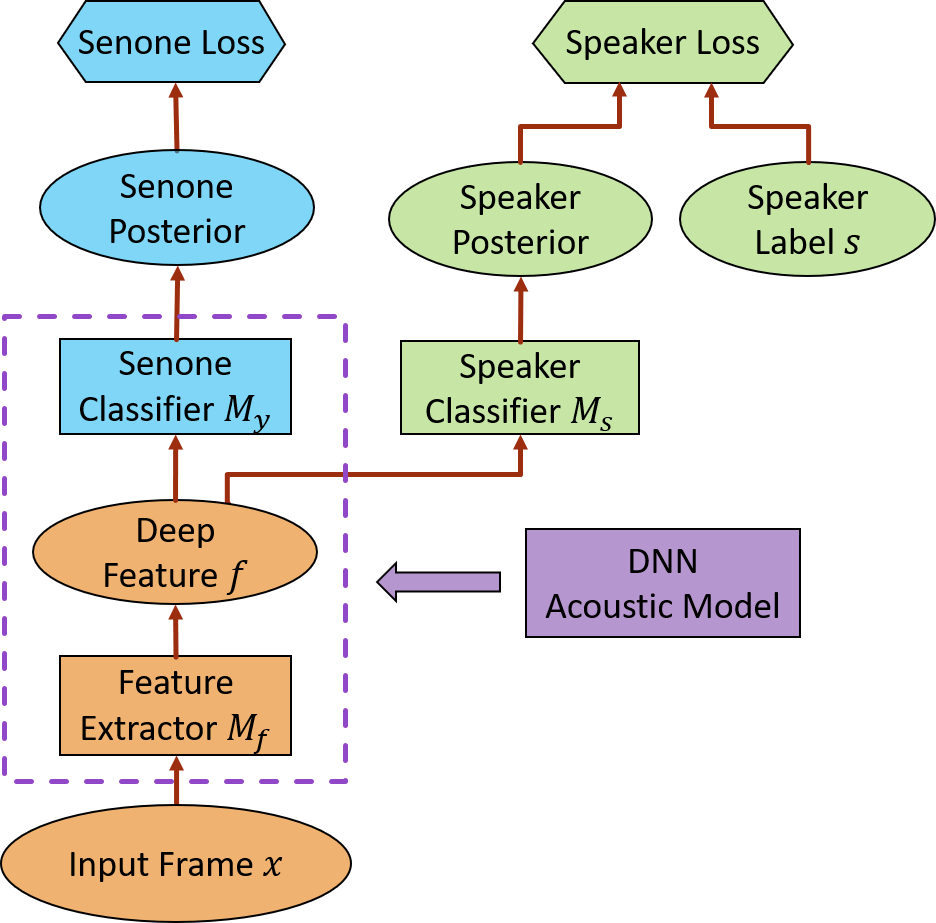}
	\caption{The framework of speaker-invariant training via adversarial learning for
	unsupervised adaptation of the acoustic models}
	\label{fig:sit}
\end{figure}
\begin{align}
	M_y(f_i) = M_y(M_f(x_i)) = p_y(q | x_i; \theta_f,
\theta_y)
	\label{eqn:senone_classify}
\end{align}
where $\mathcal{Q}$ is the set of all senones modeled by the acoustic model.

We further introduce a speaker classifier network $M_s$
which maps the deep features $F$ to the speaker posteriors $p_s(a |
f_i; \theta_s), a \in \mathcal{A}$ as follows:
\begin{align}
	M_s(M_f(x_i)) & = p_s(a | x_i; \theta_s, \theta_f)
	\label{eqn:speaker_classify}
\end{align}
where $a$ is one speaker in the set of all speakers $\mathcal{A}$. 

To make the deep features $F$ speaker-invariant, the distributions of
the features from different speakers should be as close to each other as
possible. Therefore, the $M_f$ and $M_s$ are jointly trained with an
adversarial objective, in which $\theta_f$ is adjusted to \emph{maximize}
the frame-level speaker classification loss
$\mathcal{L}_{\text{speaker}}(\theta_f, \theta_s)$ while $\theta_s$ is adjusted
to \emph{minimize} $\mathcal{L}_{\text{speaker}}(\theta_f, \theta_s)$ below:
\begin{align}
	& \mathcal{L}_{\text{speaker}}(\theta_f, \theta_s) = - \sum_{i = 1}^{N} \log
	p_s(s_i | x_i; \theta_f, \theta_s)\nonumber \\
	& \quad \quad \quad \quad \quad \quad = - \sum_{i = 1}^{N} \sum_{a\in
		\mathcal{A}} \mathbbm{1}[a =
	s_i] \log M_s(M_f(x_i)) \label{eqn:loss_cond1}
\end{align}
where $s_i$ denote the speaker label for the input frame
$x_i$. 

This minimax competition will first increase the discriminativity of 
$M_s$ and the speaker-invariance of the features generated by $M_f$, and
will eventually converge to the point where $M_f$ generates extremely
confusing features that $M_s$ is unable to distinguish.

At the same time, we want to make the deep features senone-discriminative by
minimizing the cross-entropy loss between the predicted senone posteriors
and the senone labels as follows:
\begin{align}
	\mathcal{L}_{\text{senone}}(\theta_f, \theta_y) & = -\sum_{i = 1}^N
	\log p_y(y_i|x_i;\theta_f, \theta_y) \nonumber \\
	&=-\sum_{i = 1}^N \sum_{q\in
		\mathcal{Q}} \mathbbm{1}[q =
	y_i] \log M_y(M_f(x_i))
	\label{eqn:loss_senone}
\end{align}

In SIT, the acoustic model network and the condition classifier
network are trained to jointly optimize the primary task of senone
classification and the secondary task of
speaker classification with an adversarial objective function. 
Therefore, the total loss is constructed as
\begin{align}
	&\mathcal{L}_{\text{total}}(\theta_f, \theta_y, \theta_s) =
	\mathcal{L}_{\text{senone}}(\theta_f, \theta_y) - 
	\lambda\mathcal{L}_{\text{speaker}}(\theta_s, \theta_f)
	\label{eqn:loss_total}
\end{align}
where $\lambda$ controls the trade-off between the senone loss and the speaker classification loss
 in Eq.\eqref{eqn:loss_senone} and Eq.\eqref{eqn:loss_cond1} respectively.
 
We need to find the optimal parameters $\hat{\theta}_y, \hat{\theta}_f$ and $\hat{\theta}_s$ such that
\begin{align}
    (\hat{\theta}_f, \hat{\theta}_y) = \argmin_{\theta_y, \theta_f} \mathcal{L}_{\text{total}}(\theta_f, \theta_y, \hat{\theta}_s) \\
    \hat{\theta}_s = \argmax_{\theta_s} \mathcal{L}_{\text{total}}(\hat{\theta}_f, \hat{\theta}_y, \theta_s)  
\end{align}

The parameters are updated as follows via back propagation with stochastic gradient descent (SGD):
\begin{align}
	& \theta_f \leftarrow \theta_f - \mu \left[ \frac{\partial
		\mathcal{L}_{\text{senone}}}{\partial \theta_f} - \lambda \frac{\partial
			\mathcal{L}_{\text{speaker}}}{\partial
			\theta_f}
		\right]
		\label{eqn:grad_f} \\
	& \theta_s \leftarrow \theta_s - \mu \frac{\partial
		\mathcal{L}_{\text{speaker}}}{\partial \theta_s}
		\label{eqn:grad_s} \\
	& \theta_y \leftarrow \theta_y - \mu \frac{\partial
		\mathcal{L}_{\text{senone}}}{\partial \theta_y}
	\label{eqn:grad_y}
\end{align}
where $\mu$ is the learning rate.

Note that the negative coefficient $-\lambda$ in Eq. \eqref{eqn:grad_f}
induces reversed gradient that maximizes
$\mathcal{L}_{\text{speaker}}(\theta_f, \theta_s)$ in Eq.  \eqref{eqn:loss_cond1}
and makes the deep feature speaker-invariant. 
For easy implementation, gradient reversal layer is introduced in
\cite{grl_ganin}, which acts as an identity transform in the forward propagation
and multiplies the gradient by $-\lambda$ during the backward propagation.

The optimized network consisting of $M_f$ and $M_y$ is used as the
SIT acoustic model for ASR on test data.

\section{Experiments}
\label{sec:experiment}
In this work, we perform SIT on a DNN-hidden Markov model (HMM) acoustic
model for ASR on CHiME-3 dataset.

\subsection{CHiME-3 Dataset}
The CHiME-3 dataset is released with the 3rd CHiME speech Separation and
Recognition Challenge \cite{chime3_barker}, which incorporates the Wall
Street Journal corpus sentences spoken in challenging
noisy environments, recorded using a 6-channel tablet based microphone
array.
CHiME-3 dataset consists of both real and simulated data. The real speech
data was recorded in five real noisy
environments (on buses (BUS), in caf\'{e}s (CAF), in pedestrian areas (PED), at street
junctions (STR) and in booth (BTH)). To generate the simulated data, the clean speech is first
convolved with the estimated impulse response of the environment and
then mixed with the background noise separately recorded in that
environment \cite{chime3_hori}. The noisy training data consists of 1999 real
noisy utterances from 4 speakers, and 7138 simulated noisy utterances
from 83 speakers in the WSJ0 SI-84 training set recorded in 4 noisy
environments. There are 3280 utterances in the development set including
410 real and 410 simulated utterances for each of the 4 environments.
There are 2640 utterances in the test set including 330 real and 330
simulated utterances for each of the 4 environments. The speakers in
training set, development set and the test set are mutually different
(i.e., 12 different speakers in the CHiME-3 dataset). The training,
development and test data sets are all recorded in 6 different channels. 

In the experiments, we use 9137 noisy training utterances in the CHiME-3 dataset as the
training data. The real and simulated development data
in CHiME-3 are used as the test data. Both the training and test data 
are far-field speech from the 5th microphone channel. The WSJ 5K word 
3-gram language model (LM) is used for decoding.


\subsection{Baseline System}
\label{sec:baseline}
In the baseline system, we first train an SI DNN-HMM acoustic model using 9137
noisy training utterances with cross-entropy criterion.

The 29-dimensional log Mel filterbank features together with 1st and 2nd order 
delta features (totally 87-dimensional) for both the clean and noisy
utterances are extracted by following the process in \cite{li2012improving}. Each frame is spliced together
with 5 left and 5 right context frames to form a 957-dimensional feature. The spliced
features are fed as the input of the feed-forward DNN after global mean and variance
normalization. The DNN has 7 hidden layers with 2048 hidden units for each layer. The
output layer of the DNN has 3012 output units corresponding to 3012 senone labels.
Senone-level forced alignment of the clean data is generated using a
Gaussian mixture model-HMM system. As shown in Table \ref{table:asr_wer}, the WERs for the SI DNN are 17.84\% and 17.72\%
respectively on real and simulated test data respectively. Note that our experimental setup does
not achieve the state-of-the-art performance on CHiME-3 dataset (e.g., we did not perform beamforming, sequence training or use recurrent neural network language model for decoding.) since our goal is
to simply verify the effectiveness of SIT in reducing inter-speaker variability.

\subsection{Speaker-Invariant Training for Robust Speech Recognition}
We further perform SIT on the baseline noisy DNN acoustic model with 
9137 noisy training utterances in CHiME-3. The
feature extractor $M_f$ is initialized with the first $N_h$ layers of the
DNN and the senone classifier is initialized with the rest $(7-N_h)$ hidden
layers plus the output layer. $N_h$ indicates the position of the deep
hidden feature in the acoustic model. The speaker classifier $M_s$ is a
feedforward DNN with 2 hidden layers and 512 hidden units for each layer.
The output layer of $M_s$ has 87 units predicting the posteriors of 87
speakers in the training set. $M_f$, $M_y$ and $M_s$ are jointly trained
with an adversarial multi-task objective as described in Section
\ref{sec:sit}. $N_h$ and $\lambda$ are fixed at $2$ and $3.0$ in our
experiments. The SIT DNN acoustic model achieves 16.95\% and 16.54\%
WER on the real and simulated test data respectively, which are 4.99\% and
6.66\% relative improvements over the SI DNN baseline.

\begin{table}[h]
\centering
\begin{tabular}[c]{c|c|c|c|c|c|c}
	\hline
	\hline
	System & Data & BUS & CAF & PED & STR & Avg.\\
	\hline
	\multirow{2}{*}{\begin{tabular}{@{}c@{}} SI
		\end{tabular}} & Real & 24.77 & 16.12 & 13.39 & 17.27 &
		17.84  \\
	\hhline{~------}
	& Simu & 18.07 & 21.44 & 14.68 & 16.70 & 17.72 \\
	\hline
	 \multirow{2}{*}{\begin{tabular}{@{}c@{}} SIT
		\end{tabular}} & Real & 22.91 & 15.63 & 12.77 & 16.66 &
		\textbf{16.95} \\
	\hhline{~------}
	& Simu & 16.64 & 20.23 & 13.53 & 15.96 & \textbf{16.54} \\
	\hline
	\hline
\end{tabular}
  \caption{The ASR WER (\%) performance of SI and SIT DNN acoustic models on real and simulated development set of CHiME-3.}
\label{table:asr_wer}
\end{table}

\subsection{Visualization of Deep Features}
\begin{figure}[htpb!]
	\centering
	\includegraphics[width=0.95\columnwidth]{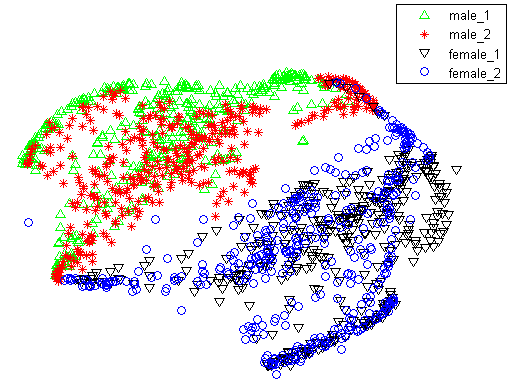}
	\caption{t-SNE visualization of the deep features $F$ generated by the SI DNN acoustic model when speech
	frames aligned with phoneme ``ah'' from two male and two female speakers in CHiME-3 training set are fed as the input.}
	\label{fig:tsne_si}
\end{figure}

\begin{figure}[htpb!]
	\centering
	\includegraphics[width=0.95\columnwidth]{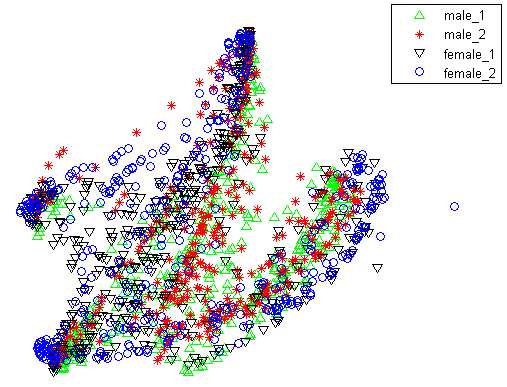}
	\caption{t-SNE visualization of the deep features $F$ generated by the SIT DNN acoustic 
	model when the same speech frames as in Fig. \ref{fig:tsne_si} are fed as the input.}
	\label{fig:tsne_sit}
\end{figure}

We randomly select two male speakers and two female speakers from the noisy training set
and extract speech frames aligned with the phoneme ``ah'' for each of the four 
speakers. In Figs. \ref{fig:tsne_si} and \ref{fig:tsne_sit}, we visualize the deep features $F$ generated by the SI
and SIT DNN acoustic models when the ``ah'' frames of the four speakers are given as the input
using t-SNE. In
Fig. \ref{fig:tsne_si}, the deep feature distributions in the SI model for the male (in red and green) and female speakers (in back and blue) are far away from each other and even the distributions for 
the speakers of the same gender are separated from each other. While after SIT, the deep
feature distributions for all the male and female speakers are well aligned with each other 
as shown in Fig. \ref{fig:tsne_sit}. The significant increase in the overlap among 
distributions of different speakers justifies that the SIT remarkably enhances the speaker-invariance of the deep features $F$. The adversarial optimization of the speaker classification loss does not just serve as a regularization term to achieve better generalization on the test data.

\subsection{Unsupervised Speaker Adaptation}
SIT aims at suppressing the effect of inter-speaker variability on DNN
acoustic model so that the acoustic model is more compact and has stronger
discriminative power. When adapted to the same test speakers, the SIT DNN
is expected to achieve higher ASR performance than the baseline SI DNN due
to the smaller overlaps among the distributions of different speech units.

In our experiment, we adapt the SI and SIT DNNs to each of the 4 speakers
in the test set in an unsupervised fashion. The constrained re-training
(CRT) \cite{erdogan2016multi} method is used for adaptation, where we
re-estimate the DNN parameters of only a subset of layers while holding the
remaining parameters fixed during cross-entropy training. The adaptation
target (1-best alignment) is obtained through the first-pass decoding of
the test data, and the second-pass decoding is performed using the
SA SI and SI DNNs. 

The WER results for unsupervised speaker
adaptation is shown in Table \ref{table:crt_wer}, in which only the bottom
2 layers of the SI and SIT DNNs are adapted during CRT. The speaker-adapted (SA)
SIT DNN achieves 15.46\% WER which is 4.86\% relatively higher than the
SA SI DNN. The CRT adaptation provides 8.91\% and 8.79\%
relative WER gains over the unadapted SI and SIT models respectively. The
lower WER after speaker adaptation indicates that SIT has effectively reduced the
high variance and overlap in an SI acoustic model caused by the
inter-speaker variability.

\begin{table}[h]
\centering
\begin{tabular}[c]{c|c|c|c|c|c}
	\hline
	\hline
	System & BUS & CAF & PED & STR & Avg.\\
	\hline
	\multirow{1}{*}{\begin{tabular}{@{}c@{}} SA SI
		\end{tabular}} & 22.76 & 15.56 & 11.52 & 15.37 &
		16.25 \\
	\hline
	 \multirow{1}{*}{\begin{tabular}{@{}c@{}} SA SIT
		\end{tabular}} & 21.42 & 14.79 & 11.11 & 14.70 &
		\textbf{15.46} \\
	\hline
	\hline
\end{tabular}
  \caption{The ASR WER (\%) performance of SA SI and SA SIT DNN
  acoustic models after CRT unsupervised speaker adaptation on real development set of CHiME-3.}
\label{table:crt_wer}
\end{table}

\section{Conclusions and Future Works}
In this work, SIT is proposed to suppress the effect of inter-speaker
variability on the SI DNN acoustic model. In SIT, a DNN acoustic model and
a speaker classifier network are jointly optimized to minimize the senone
classification loss, and simultaneously mini-maximize the speaker
classification loss. Through this adversarial multi-task learning
procedure, a feature extractor network is learned to map the input frames
from different speakers to deep hidden features that are both
\emph{speaker-invariant} and senone-discriminative. 

Evaluated on CHiME-3 dataset, the SIT DNN acoustic model achieves 4.99\% relative WER
improvement over the baseline SI DNN. With the
unsupervised adaptation towards the test speakers using CRT, the SA SIT DNN
achieves additional 8.79\% relative WER gain, which is 4.86\% relatively improved
over the SA SI DNN. With t-SNE visualization, we show that, after SIT, the deep feature distributions of different speakers are well aligned with each other, which verifies the
strong capability of SIT in reducing speaker-variability.

SIT forgoes the need of estimating any additional SI bases or speaker
representations which are necessary in other conventional approaches such as
SAT. The SIT trained DNN acoustic model can be directly used to generate the transcription 
for unseen test speakers through \emph{one-pass online} decoding. It enables a lightweight
speaker-invariant ASR system with reduced number of parameters for both training and
testing. Additional gains are achievable by performing further
unsupervised speaker adaptation on top of the SIT model.

In the future, we will evaluate the performance of the i-vector based speaker-adversarial multi-task learning \cite{saon2017english} on CHiME-3 dataset and compare it with the proposed SIT. We will perform SIT on long short-term memory-recurrent neural networks acoustic models \cite{sak2014long, meng2017deep} and compare the improvement with feedforward DNNs. Moreover, we will perform SIT on thousands of hours of data to verify its scalability to large dataset.


\vfill\pagebreak
\clearpage
{\small
\bibliography{refs}}
\bibliographystyle{IEEEbib}

\end{document}